\documentclass[twocolumn,tighten]{aastex61}

\usepackage{graphicx}
\usepackage{ulem}
\usepackage{amsmath}
\usepackage{wrapfig}

\newcommand{\obs}{{\rm{obs}}}
\newcommand{\yr}{{\rm{yr}}}
\newcommand{\Hz}{{\rm{Hz}}}
\newcommand{\GW}{{\rm{GW}}}
\newcommand{\Md}{M_{\rm{D}}}
\newcommand{\Ma}{M_{\rm{A}}}

\begin{document}
\shorttitle{Characterizing accreting DWDs with \textit{LISA} and \textit{Gaia}}
\shortauthors{Breivik and Kremer et al.}

\title{Characterizing accreting double white dwarf binaries with the \textit{Laser Interferometer Space Antenna} and \textit{Gaia}}
\email{katelyn.breivik@northwestern.edu}
\author{Katelyn Breivik}
\affil{ Department of Physics \& Astronomy, Northwestern University, Evanston, IL 60202, USA}
\affil{ Center for Interdisciplinary Exploration \& Research in Astrophysics (CIERA), Evanston, IL 60202, USA}
\author{Kyle Kremer}
\affil{ Department of Physics \& Astronomy, Northwestern University, Evanston, IL 60202, USA}
\affil{ Center for Interdisciplinary Exploration \& Research in Astrophysics (CIERA), Evanston, IL 60202, USA}
\author{Michael Bueno} 
\affil{ Department of Physics \& Astronomy, Haverford College, Haverford, PA 19041}
\author{Shane L. Larson}
\affil{ Department of Physics \& Astronomy, Northwestern University, Evanston, IL 60202, USA}
\affil{ Center for Interdisciplinary Exploration \& Research in Astrophysics (CIERA), Evanston, IL 60202, USA}
\author{Scott Coughlin}
\affil{ Center for Interdisciplinary Exploration \& Research in Astrophysics (CIERA), Evanston, IL 60202, USA}
\affil{ School of Physics \& Astronomy, Cardiff University, The Parade, Cardiff, Wales, UK, CF24 3AA}
\author{Vassiliki Kalogera}
\affil{ Department of Physics \& Astronomy, Northwestern University, Evanston, IL 60202, USA}
\affil{ Center for Interdisciplinary Exploration \& Research in Astrophysics (CIERA), Evanston, IL 60202, USA}

\begin{abstract}
We demonstrate a method to fully characterize mass-transferring double white dwarf (DWD) systems with a helium-rich (He) WD donor based on the mass--radius relationship for He WDs. Using a simulated Galactic population of DWDs, we show that donor and accretor masses can be inferred for up to $\sim\, 60$ systems observed by both \textit{LISA} and \textit{Gaia}. Half of these systems will have mass constraints $\Delta\,\Md\lesssim0.2M_{\odot}$ and $\Delta\,\Ma\lesssim2.3\,M_{\odot}$. We also show how the orbital frequency evolution due to astrophysical processes and gravitational radiation can be decoupled from the total orbital frequency evolution for up to $\sim 50$ of these systems. 
 \end{abstract}


\section{Introduction} \label{sec:intro}

Double white dwarf (DWD) binaries, which make up the most substantial fraction of close binaries in the Milky Way \citep[e.g.,][]{Marsh1995}, will be a dominant source for future space-based interferometric gravitational-wave (GW) detectors, such as the Laser Interferometer Space Antenna (\textit{LISA}; \citet{Amaro2013,Amaro2017}). As gravitational radiation drives the components of a DWD binary together, it is possible for one of the stars to fill its Roche lobe, leading to the onset of mass-transfer.

The nature of mass-transferring DWDs has been explored both observationally \citep[e.g.,][]{Warner2002,Strohmayer2004a,Strohmayer2004b,Strohmayer2005} and theoretically \citep[e.g.,][]{Kremer2017, Gokhale2007, Marsh2004}. Depending on the nature of the mass-transfer process, this can lead to the formation of an AM CVn system \citep[e.g.,][]{Nather1981,Tutukov1996,Nelemans2004} or a merger and Type Ia supernova \citep[e.g.,][]{Maoz2014,Shen2015}. AM CVn systems in which both components are WDs provide astronomers with unique ways to use GW observations in combination with electromagnetic observations to probe the physics of mass-transfer and tidal processes.

Previous studies have shown that thousands of mass-transferring DWDs are expected to be resolved by \textit{LISA} \citep[e.g.,][]{Nelemans2004,Ruiter2010,Kremer2017}.  Orbital evolution due to mass transfer or tides is expected to modify the orbital frequency evolution (chirp) from the pure gravitational radiation dominated chirp traditionally used in parameter estimation of observed systems  \citep{Valsecchi2012}. Since \textit{LISA} will observe the total orbital frequency evolution of an accreting DWD, an understanding of how astrophysically and gravitationally driven frequency evolution contribute to the total chirp is necessary to understand the system fully.

AM CVn systems with helium (He) WD donors that are undergoing stable mass-transfer are expected to follow well-constrained evolutionary tracks \citep[e.g.,][]{Nelemans2005,Deloye2007,Kalomeni2016}. This is a consequence of the mass--radius relation for He-WDs and the way the orbits of these binaries respond to mass transfer.

In the coming years, \textit{Gaia} is expected to provide distance and radial velocity measurements for more than a billion stars in the Milky Way, including a substantial population of DWD systems \citep{Carrasco2014}. Many of these systems are also expected to be observed by LISA, including several thousand detached \citep{Korol2017} or mass-transferring systems \citep{Kremer2017}. In this Letter, we explore ways that measurements made by \textit{Gaia} and \textit{LISA} can be used in conjunction with the well-constrained evolutionary tracks expected for accreting DWDs with He-WD donors to place constraints upon various orbital parameters of these systems, including the component masses. Furthermore, such observations can be used to decouple the various components of the time-derivative of the orbital frequency for these systems.

In Section\,\ref{sec:LISAdonorMass}, we discuss  well-determined evolutionary tracks for accreting DWD binaries with He-WD donors produced in our models. We discuss how these evolutionary tracks can be used to constrain the donor mass and mass-transfer rate for a particular DWD system, given an observation of the GW frequency for that system. In Section\,\ref{sec:Gaia} we explore what can be learned from DWD systems that are observed by both \textit{Gaia} and \textit{LISA}. We conclude in Section\,\ref{sec:conclusion}.

\section{Measurements of mass transfer rates and donor masses}\label{sec:LISAdonorMass}

\begin{figure}
\begin{center}
\plotone{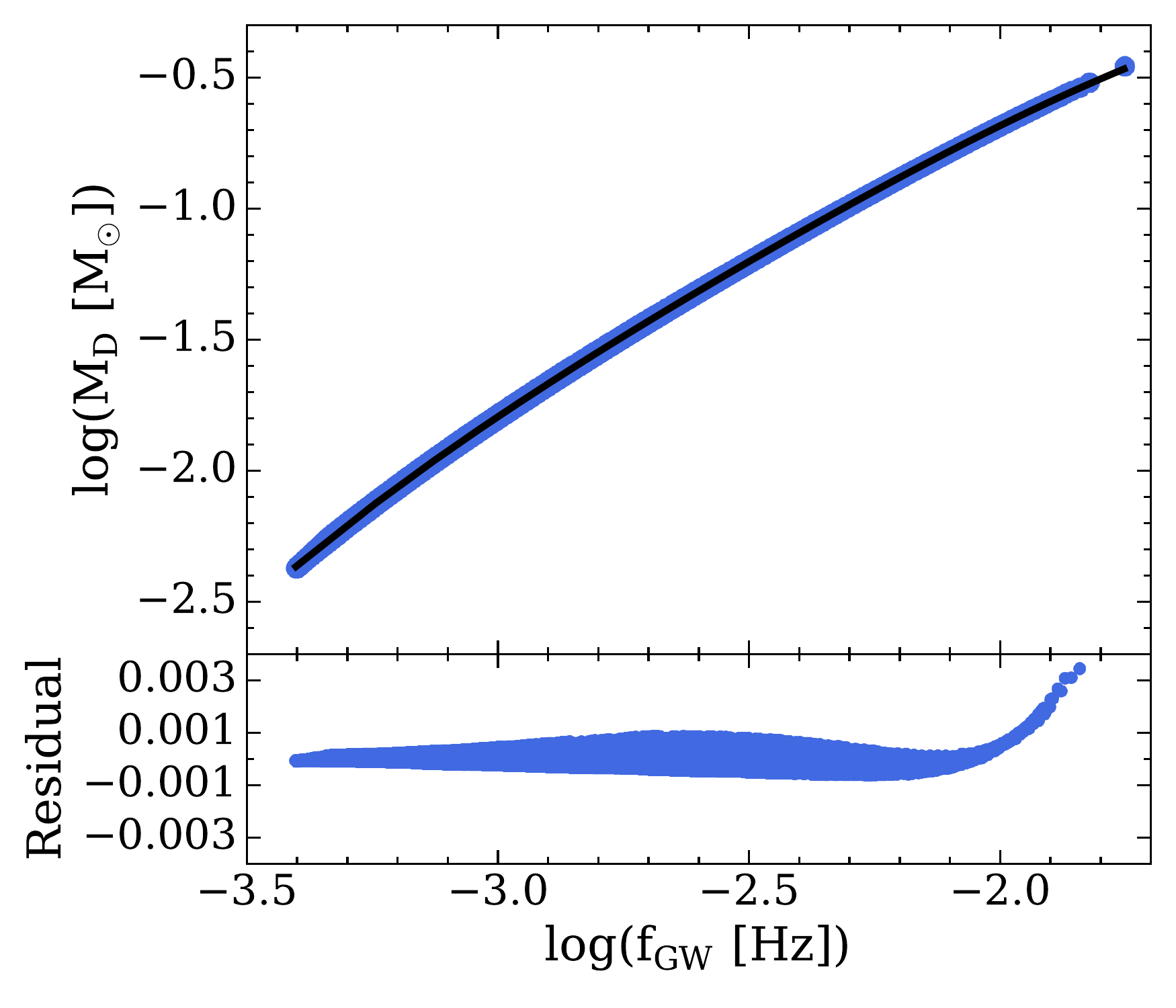}
\caption{Donor mass vs GW frequency tracks for all modeled DWD systems and the residuals ($\Md$--$M_{\rm{D,fit}}$) of the fit evaluated for each point on the tracks. \label{fig:mD_fGW}}
\end{center}
\end{figure}

Figure\,\ref{fig:mD_fGW} shows the donor mass, $\Md$, versus GW frequency, $f_{\rm{GW}}$, for the evolutionary tracks generated using the grid of initial donor and accretor masses modeled in \citet{Kremer2017}. As in previous analyses \citep[e.g.,][]{Marsh2004, Gokhale2007, Kremer2017}, we assume all mass-transferring DWDs will be circular throughout their evolution. For circular binaries, $f_{\rm{GW}} = 2/P_{\rm{orb}}$, where $P_{\rm{orb}}$ is the orbital period for the binary. The tracks shown here display the evolution of the binaries for all evolutionary time steps in which the time-derivative of the GW frequency (chirp) is negative (in the terminology of \citet{Kremer2017}, $\dot{f}_{\rm{tot}} < 0$) from the onset of mass-transfer to 10 Gyr through phases of both direct-impact and disk accretion. Each He-donor DWD that begins mass transfer will start at some point on this plot and evolve toward lower donor mass and lower GW frequency.

Recall from \citet{Kremer2017} that the initial semi-major axis for each of these systems is chosen such that the volume equivalent Roche-lobe radius of the donor is equal to the initial donor radius, as determined from the zero-temperature mass-radius (M--R) relation of \citet{Verbunt1988}.

As Figure\,\ref{fig:mD_fGW} shows, $\Md$ for all of modeled systems follow precise evolutionary tracks, independent of the initial system parameters. This narrow constraint is determined by the He-WD donor M--R relation. Similar evolutionary tracks have been shown before for accreting systems with He-WD donors \citep[e.g.,][]{Nelemans2005,Deloye2007,Kalomeni2016}.

The relation demonstrated in Figure\,\ref{fig:mD_fGW} shows that the mass-transfer rate and donor mass can be determined for mass-transferring DWDs if the GW frequency is known \textit{and} the system is observed with a measurably negative chirp, indicating a mass-transferring system. We can represent this relationship by an analytic fit to a $4\rm{th}$ order polynomial using our simulated data:

\begin{equation}
y = a+b\,x+c\,x^2+d\,x^3+e\,x^4,
\end{equation}
where $x=\log(f_{\GW}/\Hz)$, $y=\log(\Md/M_{\odot})$, $a = -2.1201$, $b=-4.2387$, $c=-3.0016$, $d=-0.7790$, and $e=-0.0791$. We choose a high order polynomial in order to closely follow the evolutionary track as well as provide a tractable method to infer donor mass for large numbers of observed systems.

We emphasize that the overall behavior of these evolutionary tracks is not unique to our mass-transfer models. As has been shown in several earlier analyses \citep[e.g.,][]{Nelemans2005,Deloye2007,Kalomeni2016}, stable mass-transfer from a He-donor WD is expected to follow similar behavior regardless of the method used to model mass-transfer. Our analysis is most similar to  \citet{Nelemans2005}. \citet{Deloye2007} use constant entropy M--R relations found in \citet{Deloye2003} and \citet{Kalomeni2016} use Eq. 3 of \citet{Nelson2003} to determine the radius as a function of mass and chemical composition. We note that systems with exceptionally precise measurements due to close proximity in the Galaxy may be used to test the M--R relation for He-WDs. Since the orbital period, donor mass and donor radius are uniquely determined by the Roche lobe, systems where the donor mass can be independently verified provide a method to infer the donor radius.

\section{Fully parameterizing the system with \textit{LISA} and \textit{Gaia}}
\label{sec:Gaia}

\begin{figure}
\plotone{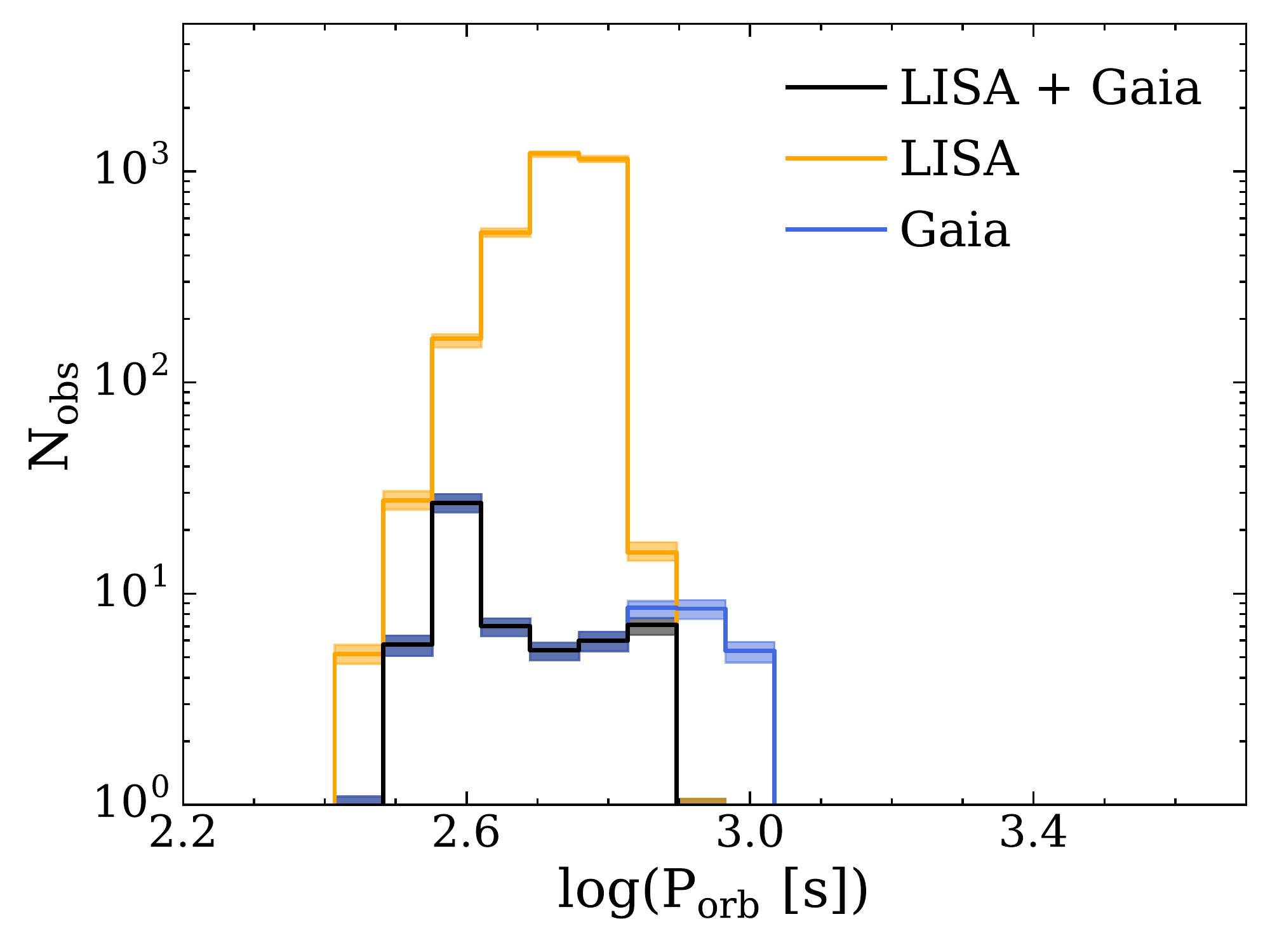}
\caption{The number of systems observed as a function of orbital period in seconds. Systems observed with negative chirps by \textit{LISA} are shown in orange. Systems with \textit{Gaia} distance measurements are shown in blue. Systems observable in both cases are shown in black. The lines show the mode of our 100 population realizations and the spread shows spread $1\sigma$ above and below this mode. \label{fig:numbers}}
\end{figure}

The donor mass for any accreting DWD can be constrained if the system is observed to be transferring mass by \textit{LISA}. If the same system is also observed by \textit{Gaia}, the chirp mass can also be constrained, allowing the accretor mass to be calculated and the orbital chirp to be decoupled into its different components. 
We use the method described in \citet{Kremer2017} to model the evolution of accreting DWD binaries with He-WD donors and build a realistic Galactic population of He-donor DWDs in the disk and bulge at the present day.  

In addition to the models used in \citet{Kremer2017}, we include five new models based on the results of \citet{Zorotovic2010,Toonen2013,Camacho2014} which suggest that the common envelope efficiency for WD-main sequence systems should be low ($\alpha\lesssim0.2$). We also include models which compute the binding energy of the stellar envelope ($\lambda$) based on the stellar parameters and stellar type according to \citet{Xu2010}. Each of our models is listed in Table\,\ref{tbl:Nobs}. We take $\alpha=0.25$, $\lambda=\rm{Var}$ to be our fiducial model.

Our methods used to estimate the \textit{Gaia}-detectability, including optical emission and extinction, are described in detail in the Appendix.  We note that the methods demonstrated can be applied to any simulated data sets with He-WD donors due to the expected behavior from the M--R relationship for He-WDs.

We generate $100$ population realizations for the Milky Way disk and bulge using the Compact Object Synthesis and Monte-Carlo Code (\texttt{COSMIC}) according to the methods described in \citet{Kremer2017}. We use these populations to investigate the overall number and binary parameters of accreting He-donor DWDs observable by both \textit{LISA} and \textit{Gaia}. Figure\,\ref{fig:numbers} shows the number and orbital period distribution of He-donor DWDs observed with negative chirps by \textit{LISA} (green) and with measured distances by \textit{Gaia} (blue) as well as systems observed by both missions (black) as a function of orbital period. From our $100$ population realizations for the fiducial model, we find, on average, $\sim3000$ systems resolved with negative chirps by \textit{LISA}, $\sim80$ systems observed by \textit{Gaia}, and $\sim 60$  systems with resolved negative chirps \emph{and} measured distances. The numbers for each of the common envelope models in \citet{Kremer2017} as well as our newly run models are summarized in Table\,\ref{tbl:Nobs}. Broadly, increasing the common envelope efficiency leads to higher numbers of observed systems.

\begin{table}[t!]
\begin{center}
\caption{Average number of He-donor DWDs observed with negative chirps by \textit{LISA} ($N_{\textit{LISA}}$), measured distances by \textit{Gaia} ($N_{\textit{Gaia}}$), and systems satisfying both conditions ($N_{\rm{both}}$). The fiducial model is $\alpha=0.25$, $\lambda=\rm{Var}$. \label{tbl:Nobs}}
\begin{tabular}{cccccc}
\tableline
\tableline
Model &  $N_{\textit{LISA}}$ &  $N_{\textit{Gaia}}$ & $N_{\rm{both}}$\\
\tableline
$\alpha=0.25$, $\lambda=\rm{Var}$ & 3077 &  78 & 61\\
$\alpha=0.25$, $\lambda=1.0$ & 3410 & 114 & 94\\
\tableline
$\alpha=0.5$, $\lambda=\rm{Var}$ & 3917 &  116 & 95\\
$\alpha=0.5$, $\lambda=1.0$ & 2757 & 73 & 60\\
\tableline
$\alpha=1.0$, $\lambda=\rm{Var}$ & 8295 &  84 & 9\\
$\alpha=1.0$, $\lambda=0.1$ & 77 & 209 & 8\\
$\alpha=1.0$, $\lambda=1.0$ & 6225 &  2305 & 518\\
$\alpha=1.0$, $\lambda=10.0$ & 3684 &  1488 & 256\\
\tableline
\tableline
\end{tabular}
\end{center}
\end{table}

\subsection{Decoupling the component masses}
\label{sec:decoupling}
\begin{figure}
\plotone{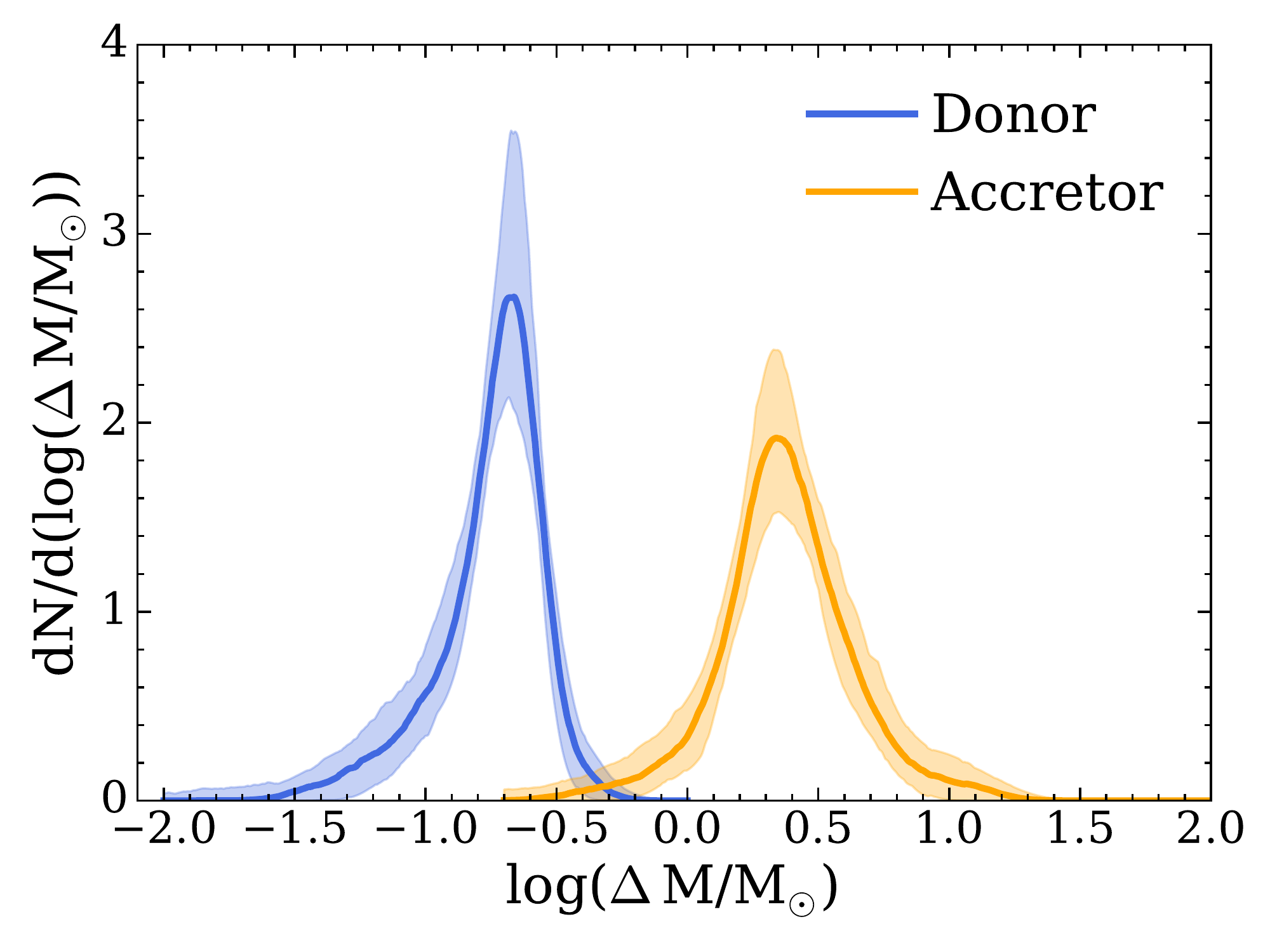}
\caption{Measurement errors for donor masses (blue) and accretor masses (orange). The shaded regions show $5-95\%$ percentile spread for our $100$ population realizations.\label{fig:errors}}
\end{figure}

For circular binaries, the signal-to-noise ratio ($\rm{S/N}$) is given by
\begin{equation}
\label{SNR}
\rm{S/N} \approx \frac{h_0 \sqrt{T_{obs}}}{h_f},
\end{equation}
where $h_0$ is the scaling amplitude,
\begin{equation}
\label{eq:h_0}
h_{\rm{0}} = 4\frac{G}{c^2}\frac{M_c}{D}\left(\frac{G}{c^3}\pi f_{\GW} M_c \right)^{2/3},
\end{equation}
$T_{\rm{obs}}$ is the observational time of the \textit{LISA} mission, taken here, as in \citet{Kremer2017} to be 4 years, $h_{\rm{f}}$ is the spectral amplitude value for a specified gravitational-wave frequency, $f_{\GW}$, given by the standard \textit{LISA} sensitivity curve in \cite{Larson2002}, and $M_c$ is the chirp mass defined as

\begin{equation}
\label{chirp}
M_c = \frac{(\Ma \Md)^{3/5}}{(\Ma + \Md)^{1/5}}.
\end{equation}
A parallax distance measurement, $D$, obtained from \textit{Gaia} can be combined with the \textit{LISA} observation of $h_0$ and $f_{\GW}$ to directly compute $M_c$ from equation \ref{eq:h_0}. Using the donor mass determined with the method of Section\,\ref{sec:LISAdonorMass}, the accretor mass can be constrained.

We assume LISA measurement errors using Eqs. 12-14 of \citet{Takahashi2002}, which are valid for our data set containing GW frequencies $10^{-4}\,\Hz<f_{GW}<10^{-2}\,\Hz$. We list them here for convenience:

\begin{align}
\label{eq:GWerrors}
\frac{\Delta\,h_0}{h_0} &= 0.2\,\Big(\frac{S/N}{10}\Big)^{-1}\\
\Delta\,f_{\GW} &= 0.22\,\Big(\frac{S/N}{10}\Big)^{-1}\Big(\frac{T_{\obs}}{\yr}\Big)^{-1}\\
\Delta\,\dot{f}_{\rm{tot}} &= 0.43\,\Big(\frac{S/N}{10}\Big)^{-1}\Big(\frac{T_{\obs}}{\yr}\Big)^{-2}.\\
\end{align}
We assume the \textit{Gaia} distance measurement error to be
\begin{equation}
\label{eq:DistErrors}
\frac{\Delta\,D}{D} = \frac{\Delta\,\alpha}{\alpha}
\end{equation}
where the distance is in $\rm{pc}$ and $\alpha$ is the \textit{Gaia} magnitude-dependent astrometric accuracy in $\rm{arcsec}$ taken from \citet{Gaia2016b}. These measurement errors can be propagated through our equations for $h_0$, $M_{c}$ and $\Md$ to obtain measurement errors for both component masses.

We note that systems with parallax measurement errors in excess of $20\%$ (approximately half of our resolved systems) will not follow this simple relation \citep{Bailer-Jones2015}. For the purposes of this initial study, we use the approximation of Eq.\,\ref{eq:DistErrors} and leave a more detailed treatment of distance errors for a later study.

Figure\,\ref{fig:errors} shows the distribution of our mass measurement errors for both $M_D$ and $M_A$ as well as the percent error between the mean `inferred' masses from observations and our $\Md - f_{\GW}$ fit  and the `true' simulated values. The peak in the $\Delta \Md$ distribution is $\sim0.2\,M_{\odot}$ while the peak for the $\Delta \Ma$ distribution is $\sim2M_{\odot}$. The $50$th percentile of our data ($\sim 30$ systems) have mass measurements better than $\Delta \Md \simeq 0.2\, M_{\odot}$ and $\Delta \Ma \simeq 2.3\, M_{\odot}$. 

\subsection{Decoupling the astrophysical chirp from the gravitation-radiation chirp}
\label{sec:chirp}
\begin{figure}[t]
\plotone{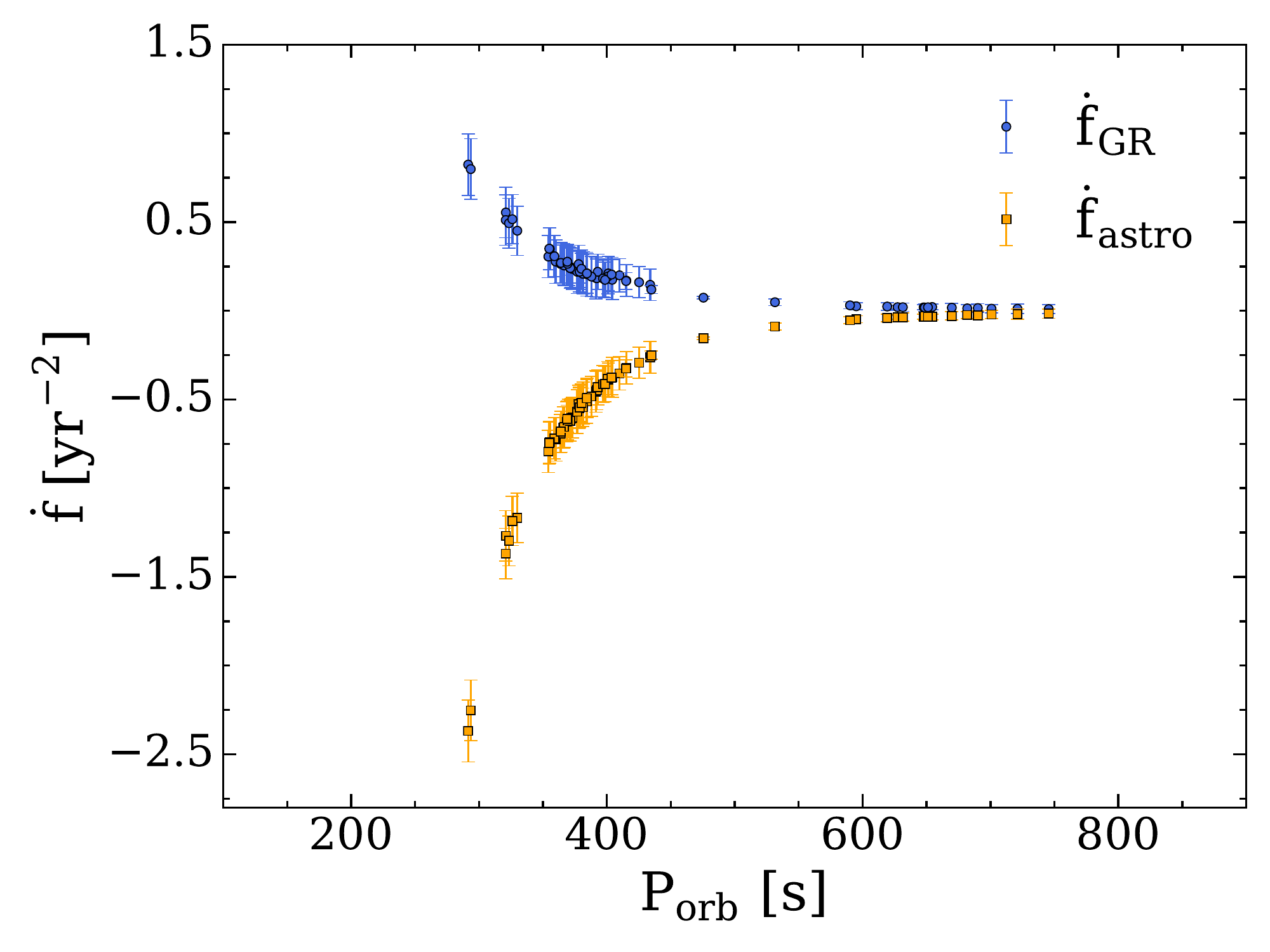}
\caption{GR (blue) and astrophysical (orange) chirps as a function of GW frequency. The Error bars show the $1\sigma$ measurement errors. See Eqs.\,\ref{eq:fdotTot}--\ref{eq:fdot_astro} for an explanation of the difference between $\dot{f}_{\rm{GR}}$ and $\dot{f}_{\rm{astro}}$.\label{fig:chirps}}
\end{figure} 

As in \citet{Kremer2017}, the chirp for mass-transferring DWDs can be broken into three separate components

\begin{equation}
\label{eq:fdotTot}\dot{f}_{\rm{total}} = \dot f_{\rm{GR}} + \dot f_{\rm{MT}}  + \dot f_{\rm{tides}}
\end{equation}
where $\dot{f}_{\rm{GR}}$, $\dot{f}_{\rm{MT}}$, and $\dot{f}_{\rm{tides}}$ are the contributions to the chirp due to gravitational radiation, mass transfer, and tidal interactions, respectively. It is convenient to group $\dot{f}_{\rm{MT}}$ and $\dot{f}_{\rm{tides}}$ into a single quantity defined as the astrophysical chirp, $\dot{f}_{\rm{astro}}$, so that $\dot{f}_{\rm{total}} = \dot f_{\rm{GR}} + \dot f_{\rm{astro}}$.

Here $\dot{f}_{\rm{GR}}$ is given by
\begin{equation}
\label{eq:fdot_GR}
\dot{f}_{\rm{GR}} = \frac{96}{5} \frac{f_{\GW}^2}{\pi c^5} \Big( G \, \pi \, f_{\GW} \, M_c \Big)^{5/3}.
\end{equation}
If $M_c$ is computed using $D$ obtained from a \textit{Gaia} observation, as described in Section\,\ref{sec:decoupling}, $\dot{f}_{\rm{GR}}$ can be computed directly. If, additionally, the system has a sufficiently high $\dot{f}_{\rm{total}}$ to be measured directly by \textit{LISA} (for $T_{\rm{obs}} = 4$ years, $\dot{f}_{\rm{min}} = 7.93 \times 10^{-10} \rm{Hz}\,\rm{yr}^{-1}$; see \citet{Kremer2017} for further detail), the observed $\dot{f}_{\rm{total}}$ and calculated $\dot{f}_{\rm{GR}}$ values can be used to solve for the astrophysical contribution to the total chirp
\begin{equation}
\dot{f}_{\rm{astro}} = \dot{f}_{\rm{total}} - \dot{f}_{\rm{GR}}.
\label{eq:fdot_astro}
\end{equation}

We propagate the measurement errors  $\Delta\,h_0$, $\Delta\,f_{\GW}$, $\Delta\,\dot{f}_{\rm{total}}$, and $\Delta\,D$ through Eqs.\,\ref{eq:fdot_GR} and \ref{eq:fdot_astro} to obtain the inferred values of $\dot{f}_{\GW}$ and $\dot{f}_{\rm{astro}}$. Figure\,\ref{fig:chirps} shows $\dot{f}_{\rm{GR}}$ (blue) and $\dot{f}_{\rm{astro}}$ (green) and their $1\sigma$ measurement error for the \textit{LISA}-resolved systems taken from a single sample realization for our fiducial model. On average, we find $\sim 50$ systems with $P_{\rm{orb}}\lesssim800\,s$ have resolvable gravitational radiation and astrophysically driven chirps from our Galactic realizations. 

\section{Conclusion} \label{sec:conclusion}
We have explored the evolution of accreting DWD binaries with He-WD donors and demonstrated these systems provide unique laboratories for probing the physics of mass-transfer in binary systems. We have also shown that if the GW frequency for these systems is obtained using \textit{LISA}, the donor mass can be constrained using well-determined evolutionary tracks. The method demonstrated above presents a previously undescribed analysis for inferring donor and accretor masses using multi-messenger observations of accreting DWDs. We note that if the systems are eclipsing, the binary may be fully parameterized by solving the `visual binary' problem. However, this requires a narrow range of orbital inclinations that will greatly reduce the overall number of characterized systems.

Furthermore, we have demonstrated that of the several thousand accreting DWDs observable by \textit{LISA}, $\sim 680$ are expected to also be observed by \textit{Gaia}. For systems observed by both \textit{LISA} and \textit{Gaia}, we have shown that in addition to the donor mass, the chirp mass, and therefore the accretor mass, can also be constrained to $\Delta\,\Md\lesssim\,0.2M_{\odot}$ and $\Delta\,\Ma\lesssim\,2.3M_{\odot}$ accuracy for $\sim30$ He-donor DWDs. Additionally, the chirp for $\sim 50$ systems can be decoupled into its GW and astrophysical components. These numbers vary if different binary evolution models are used (see Table\,\ref{tbl:Nobs}), however the methods of inferring binary parameters are agnostic to these models. Thus, while the overall number of observed accreting He-donor DWDs may change, the ability to constrain the donor mass, accretor mass, and orbital frequency evolution remains.   

The method demonstrated above relies on an understanding of the mass--radius relationship that governs He-WDs. While several models currently exist, future observations of mass-transferring DWDs with He-WD donors are needed to properly constrain the physics. Current and future observing campaigns like the \textit{Gaia} mission and the Large Synoptic Survey Telescope \citep{LSST2009} show promise to discover such systems and, through follow-up observations, constrain the He-WD mass--radius relation before \textit{LISA}'s launch. 

\acknowledgements

KB gratefully acknowledges Chris Pankow's helpful discussion. KB and KK also gratefully acknowledge Michael Zevin for helpful discussion. KK acknowledges support from the National Science Foundation Graduate Research Fellowship Program under Grant No. DGE-1324585. KB and SLL acknowledge support from NASA Grant NNX13AM10G. VK acknowledges support from Northwestern University. The majority of our analysis was performed using the computational resources of the Quest high performance computing facility at Northwestern University which is jointly supported by the Office of the Provost, the Office for Research, and Northwestern University Information Technology.

\listofchanges

\appendix

\section{Modeling the optical emission}

Here we introduce our method for modeling the electromagnetic emission of accreting DWD systems. We explore optical emission from the component stars and the disk itself.

\subsection{Emission from WD components}
As in \citet{Nelemans2004}, we consider three sources of optical emission: (1) the donor, (2) the accretor, and (3) the accretion disk (if present). The optical emission from the donor and accretor is modeled as the cooling luminosity of the WD. We use the cooling functions of \citet{Nelemans2004}, which are approximations to those of \citet{Hansen1999}:

\begin{equation}
\log L = L_{\rm{max}} - 1.33\log \left( \frac{t}{10^6\,\rm{yrs}} \right)
\end{equation}
where $L_{\rm{max}}$ is given by:
\begin{equation}
L_{\rm{max}} = 
\begin{cases}
	1-(0.9-M_{\rm{WD}}), & M_{\rm{WD}} \geq 0.5 M_{\odot}\\
	1.4-1.33(0.45-M_{\rm{WD}}), & M_{\rm{WD}} < 0.5 M_{\odot} 
\end{cases}
\end{equation}

In this simple analysis, as in \citet{Nelemans2004}, we do not consider heating of the accretor due to accretion.

Assuming, the WDs radiate as blackbodies, the temperature of each star is given by $L = 4 \pi R_{\rm{WD}}^2  \sigma T^4$. From the temperature and appropriate bolometric correction, the optical emission from each component can be calculated. 

\subsection{Emission from disk}
The accretion luminosity for a mass-transferring binary is given by
\begin{equation}
L_{\rm{acc}} = G M_A \dot{M} \left( \frac{1}{R_A} - \frac{1}{R_{\rm{L1}}} \right)
\end{equation}
If an accretion disc is present, we assume half of the accretion luminosity is radiated by the disk itself giving
\begin{equation}
L_{\rm{disk}} = \frac{1}{2}G M_A \dot{M} \left( \frac{1}{R_A} - \frac{1}{R_{\rm{L1}}} \right),
\end{equation}
with the other half being radiated at the boundary layer. Here, $R_A$ is the accretor radius, $\dot{M}$ is the mass-transfer rate, and $R_{\rm{L1}}$ is the distance of the first Lagrangian point to the center of the accretor. For circular binaries, $R_{\rm{L1}} = a - R_{\rm{L}}$, where $R_{\rm{L}}$ is the Roche-lobe radius as determined by \citet{Eggleton1983}.

We assume the disk has a radial temperature profile,\citep{Pringle1981}
\begin{equation}
T(R) = \Big(\frac{3GM_A\dot{M}}{8\pi r^3 \sigma}[1-(R_A/R)^{1/2}]\Big)^{1/4},
\end{equation}
and is made up of $10$ equally-radially-spaced annuli, each radiating as a blackbody. As in \citet{Nelemans2001}, we take the outer radius of the disk to be $R_{\rm{out}}=0.7 R_{R_{\rm{L1}}}$. We compute visual magnitudes, $m_V$, and $B-V$ colors from the  bolometric corrections in \citet{Flower1996} with corrections of \citet{Torres2010} using the blackbody luminosity and effective temperature of the disk. The \textit{Gaia} $G$-magnitude is then computed from the color--color transformations of \citet{Jordi2010}. 

For the sake of comparison, we also take interstellar reddening into account using the simple \citet{Sandage1972} extinction model used in \citet{Nelemans2004}:
\begin{equation}\label{eq:extinct} A_{V}(\infty)=\left\{
    \begin{array}{@{} l c @{}}
      0.165 \frac{[\tan(50^{\circ})-\tan(b)]}{\sin(b)} & \text{for }b <  50^{\circ} \\
      0 &  \text{for }b \geq  50^{\circ},
    \end{array}\right.
\end{equation}
where $b$ is the galactic latitude. To apply this extinction model to our DWD populations, we account for the extinction of each population by computing the integrated spatial distribution of each galactic component (disk, bulge) and multiply by the total extinction. For our disk population, we compute $A_V(d)_{disk}$ as 
\begin{equation}\label{eq:AVtd}
A_V(d)_{\rm{disk}} = A_V(\infty)\ \tanh\Big(\frac{d\ \sin(b)}{z_h}\Big),
\end{equation}
where $d$ is the distance to each DWD and $z_h=0.352\ \rm{kpc}$. For the bulge, we compute $A_V(d)_{\rm{bulge}}$ as
\begin{equation}
A_V(d)_{\rm{bulge}} = A_V(\infty)\ Erf\Big(\frac{d\sin(b)}{R_h}\Big),
\end{equation}
where $R_h = 0.5\ \rm{kpc}$. 

We use the color transformations in \cite{Cardelli1989} to convert the $V$-band extinction to $G$-band extinction using the center of the \textit{Gaia} wavelength band $\lambda_G = 673\ \rm{nm}$. Here we also note that the maximal extinction is $A_V\leq0.2$ which gives $A_{\lambda_G} \leq 0.16$. Though our extinction model is simplified, we note that the overall effects from reddening on the number of DWDs detectable by \textit{Gaia} are outweighed by the differences of our binary evolution models. Thus we do not expect that using a different extinction model will have a significant effect on our end results.


\vspace{5mm}


\begin{thebibliography}{}
\bibitem[Amaro-Seoane et al.\,(2013)] {Amaro2013} Amaro-Seoane, P. et al. 2013, arXiv: 1305.5720
\bibitem[Amaro-Seoane et al.\,(2017)] {Amaro2017} Amaro-Seoane, P. et al. 2017, arXiv: 1702.00786
\bibitem[Bailer-Jones(2015)]{Bailer-Jones2015} Bailer-Jones, C.~A.~L.\ 2015, \pasp, 127, 994 
\bibitem[Camacho et al.(2014)]{Camacho2014} Camacho, J., Torres, S., Garc{\'{\i}}a-Berro, E., et al.\ 2014, \aap, 566, A86 
\bibitem[Cardelli et al.(1989)]{Cardelli1989} Cardelli, J.~A., Clayton, G.~C., \& Mathis, J.~S.\ 1989, \apj, 345, 245 
\bibitem[Carrasco et al.(2014)]{Carrasco2014} Carrasco, J.~M., Catal{\'a}n, S., Jordi, C., et al.\ 2014, \aap, 565, A11
\bibitem[Deloye \& Bildsten(2003)]{Deloye2003} Deloye, C.~J., \& Bildsten, L.\ 2003, \apj, 598, 1217 
\bibitem [Deloye et al.$\,$(2007)] {Deloye2007} Deloye, C. J., Taam, R. E., Winisdoerffer, C., \& Chabrier, G. MNRAS, 381, 525
\bibitem[Eggleton$\,$(1983)] {Eggleton1983} Eggleton, P. P. 1983, ApJ, 268, 368.
\bibitem[Flower(1996)]{Flower1996} Flower, P.~J.\ 1996, \apj, 469, 355 
\bibitem[Gaia Collaboration et al.(2016b)]{Gaia2016b} Gaia Collaboration, Brown, A.~G.~A., Vallenari, A., et al.\ 2016, \aap, 595, A2
\bibitem [Gokhale et al.$\,$(2007)] {Gokhale2007} Gokhale, V., Peng, X. M. and Frank, J. 2007, ApJ 655, 1010.
\bibitem [Hansen$\,$(1999)] {Hansen1999} Hansen, B. M. S. 1999, ApJ, 520, 680
\bibitem[Hurley et al.$\,$(2002)]{Hurley2002} Hurley, J. R., Tout, C. A., \& Pols, O. R. 2002, MNRAS, 329, 897.
\bibitem[Jordi et al.(2010)]{Jordi2010} Jordi, C., Gebran, M., Carrasco, J.~M., et al.\ 2010, \aap, 523, A48 
\bibitem [Kalomeni et al.$\,$(2016)]{Kalomeni2016} Kalomeni, B., Nelson, L., Rappaport, S. et al. 2016, \apj, 833, 83
\bibitem[Korol et al.(2017)]{Korol2017} Korol, V., Rossi, E.~M., Groot, P.~J., et al.\ 2017, \mnras, 470, 1894 
\bibitem [Kremer et al.$\,$(2017)] {Kremer2017} Kremer, K., Breivik, K., Larson, S. L., \& Kalogera, V. 2017, \apj, 846, 95
\bibitem[Larson et al.(2002)]{Larson2002} Larson, S.~L., Wellings, R.~W., \&Hiscock, W.~A., \prd, 66, 062001 
\bibitem[LSST Science Collaboration et al.(2009)]{LSST2009} LSST Science Collaboration, Abell, P.~A., Allison, J., et al.\ 2009, arXiv:0912.0201
\bibitem [Maoz et al.$\,$(2014)] {Maoz2014} Maoz, D., Mannucci, F., and Nelemans, G. 2014, AARA, 52, 107.
\bibitem[Marsh et al.$\,$(1995)] {Marsh1995} Marsh, T. R., Dhillon, V. S. and Duck, S. R. 1995, MNRAS, 275, 828.
\bibitem [Marsh et al.$\,$(2004)] {Marsh2004} Marsh, T. R., Nelemans, G. and Steeghs, D. 2004, MNRAS, 350, 113.
\bibitem [Nather et al.$\,$(1981)] {Nather1981} Nather, R. E., Robinson, E. L. and Stover, R. J. 1981, ApJ, 244, 269.
\bibitem [Nelemans et al.$\,$(2001)] {Nelemans2001} Nelemans, G., Portegies Zwart, S. F., Verbunt, F., \& Yungelson, L. R.  2001, A\&A, 368, 939
\bibitem [Nelemans et al.$\,$(2004)] {Nelemans2004} Nelemans, G., Yungelson, L. R., \& Portegies Zwart, S. F. 2004, MNRAS, 349, 181
\bibitem [Nelemans$\,$(2005)] {Nelemans2005} Nelemans, G. The Astrophysics of Cataclysmic Variables and Related Objects, Proceedings of ASP Conference Vol. 330. Edited by J.-M. Hameury and J.-P. Lasota. San Francisco: Astronomical Society of the Pacific, 2005., p.27
\bibitem[Nelson \& Rappaport(2003)]{Nelson2003} Nelson, L.~A., \& Rappaport, S.\ 2003, \apj, 598, 431 
\bibitem [Pringle$\,$(1977)] {Pringle1977} Pringle, J. E. 1977, MNRAS, 178, 195
\bibitem[Pringle(1981)]{Pringle1981} Pringle, J.~E.\ 1981, \araa, 19, 137 
\bibitem[Ruiter et al.(2010)]{Ruiter2010} Ruiter, A.~J., Belczynski, K., Benacquista, M., Larson, S.~L., \& Williams, G.\ 2010, \apj, 717, 1006 
\bibitem[Sandage(1972)]{Sandage1972} Sandage, A.\ 1972, \apj, 178, 1 
\bibitem[Shen$\,$(2015)] {Shen2015} Shen, K. 2015, ApJL,805, L6.
\bibitem[Strohmayer$\,$(2004a)] {Strohmayer2004a} Strohmayer, T. 2004a, ApJ 610, 416.
\bibitem[Strohmayer$\,$(2004b)] {Strohmayer2004b} Strohmayer, T. 2004b, ApJ 614, 358.
\bibitem[Strohmayer$\,$(2005)] {Strohmayer2005} Strohmayer, T. 2005, ApJ, 627, 920
\bibitem[Takahashi \& Seto(2002)]{Takahashi2002} Takahashi, R., \& Seto, N.\ 2002, \apj, 575, 1030 
\bibitem[Torres(2010)]{Torres2010} Torres, G.\ 2010, \aj, 140, 1158 
\bibitem[Toonen \& Nelemans(2013)]{Toonen2013} Toonen, S., \& Nelemans, G.\ 2013, \aap, 557, A87 
\bibitem[Tutukov and Yungelson$\,$(1996)] {Tutukov1996} Tutukov, A. and Yungelson, L. 1996, MNRAS, 280, 1035.
\bibitem[Valsecchi et al.(2012)]{Valsecchi2012} Valsecchi, F., Farr, W.~M., Willems, B., Deloye, C.~J., \& Kalogera, V.\ 2012, \apj, 745, 137 
\bibitem [Verbunt and Rappaport$\,$(1988)] {Verbunt1988} Verbunt, F. and Rappaport, S. 1988, ApJ, 332, 193
\bibitem [Wade et al.$\,$(1984)] {Wade1984} Wade, R. A. 1984, MNRAS, 208, 381
\bibitem [Warner and Woudt$\,$(2002)] {Warner2002} Warner, B. and Woudt, P. 2002, PASP 792, 129
\bibitem[Xu \& Li(2010)]{Xu2010} Xu, X.-J., \& Li, X.-D.\ 2010, \apj, 716, 114
\bibitem[Zorotovic et al.(2010)]{Zorotovic2010} Zorotovic, M., Schreiber, M.~R., G{\"a}nsicke, B.~T., \& Nebot G{\'o}mez-Mor{\'a}n, A.\ 2010, \aap, 520, A86 



\end{thebibliography}
\end{document}